\documentclass[draftclsnofoot,onecolumn,12pt]{IEEEtran}
\pdfoutput=1
\usepackage{float}
\usepackage{amsmath}        
\usepackage{graphics}       
\usepackage{longtable}          
\usepackage{setspace}
\usepackage{amssymb}
\usepackage{lscape}
\usepackage[noadjust]{cite}
\usepackage{algorithm}
\usepackage{algorithmic}
\usepackage{epsfig}
\usepackage{color}
\usepackage[update,prepend]{epstopdf}
\usepackage{microtype}

\begin{document}
\title{A Sub-Space Method to Detect Multiple Wireless Microphone Signals in TV Band White Space}

\author{Harpreet S. Dhillon, Jeong-O Jeong, Dinesh Datla, Michael Benonis, R. Michael Buehrer and Jeffrey H. Reed
\thanks{
Harpreet S. Dhillon is with the Wireless Networking and Communications Group (WNCG), Department of Electrical and Computer Engineering, The University of Texas at Austin, 1 University Station C0803, Austin, TX 78712 (Email: harpreet@mail.utexas.edu). Jeong-O Jeong, Dinesh Datla, Michael Benonis, R. Michael Buehrer and Jeffrey H. Reed are with the Wireless@Virginia Tech, Department of Electrical and Computer Engineering, Virginia Polytechnic Institute and State University, Blacksburg, VA-24061 (Email: \{jeongo9, ddatla, mjb8h, buehrer, reedjh \}@vt.edu).
Manuscript last revised: \today.}
}

\maketitle
\begin{abstract}
The main hurdle in the realization of dynamic spectrum access (DSA) systems from physical layer perspective is the reliable sensing of low power licensed users. One such scenario shows up in the unlicensed use of TV bands where the TV Band Devices (TVBDs) are required to sense extremely low power wireless microphones (WMs). The lack of technical standard among various wireless manufacturers and the resemblance of certain WM signals to narrow-band interference signals, such as spurious emissions, further aggravate the problem. Due to these uncertainties, it is extremely difficult to abstract the features of WM signals and hence develop robust sensing algorithms. To partly counter these challenges, we develop a two-stage sub-space algorithm that detects multiple narrow-band analog frequency-modulated signals generated by WMs. The performance of the algorithm is verified by using experimentally captured low power WM signals with received power ranging from -100 to -105 dBm. 
The problem of differentiating between the WM and other narrow-band signals is left as a future work.
\end{abstract}


%
%
\section{Introduction}
Classical solution to the problem of spectrum allocation has been to divide it into non-overlapping frequency bands and assign each band to a wireless user/technology. While this avoids interference between users, it does not necessarily result in efficient spectrum utilization~\cite{FCC02}. This triggered a lot of interest in the wireless communications community to develop efficient methods that dynamically utilize the unused spectrum~\cite{Mit99, ZhaSad07, AkyLee06, Hay05}. In particular, the significant white space present in the TV bands (VHF/UHF) received a considerable attention and led to the conception of IEEE $802.22$ wireless regional area networks (WRAN)~\cite{Commission2002,Commission2004}. The basic underlying idea is to allow unlicensed users to use the licensed spectrum without causing harmful interference to the licensed users. One of the ways for the unlicensed users to achieve this is to identify the white spaces by spectrum sensing and then evacuate the spectrum when licensed user tries to access it. While spectrum sensing was a requirement for unlicensed TV Band Devices (TVBDs) in the FCC's Second Report and Order issued in 2008~\cite{Commission2008}, devices which access a TV Band Database are no longer required by law to perform spectrum sensing~\cite{Commission2010}. In stead, they are legally required to only perform geolocation in order to avoid causing harmful interference to licensed users of the spectrum. This is perhaps due to the limited performance of the sensing prototypes received by the FCC, and hence the need of efficient spectrum sensing still remains central to the success of decentralized DSA.

The cognitive use of the TV Band requires protecting licensed ATSC, NTSC, and Broadcast Auxiliary Service (BAS) devices including Wireless Microphones (WMs). While there has been significant progress in developing detection methods for TV signals due to their defined structure, it has not been the case for WMs~\cite{CheGaoDau07, She08, SteCho09}. The main challenges has been the low transmit power of the WMs, lack of technical standard among various manufacturers and the resemblance of certain WM signals to narrow-band interference signals, such as spurious emissions~\cite{SheSadZha09}. WMs typically operate with a transmit power ranging from 10 mW to 250 mW, though typical professional WMs operate with a transmit power of around 50 mW. Since WMs are used by live performers, their location is frequently in motion, and hence the field strength at any given point in space varies rapidly as a function of time due to fading and multipath interference.  Almost all WMs use analog Frequency Modulation (FM), with a channel bandwidth of no greater than 200 kHz~\cite{Commission2010B}.  WMs may also operate on any carrier frequency that is a multiple of 25 kHz, and 25 kHz from the edge of the TV bands. Along with these uncertainties, WM signal detection is further complicated by real-world transmitter designs. Because of their relatively narrowband nature, nonlinear active devices such as amplifiers can generate Intermodulation Distortion (IMD). Such signals may be generated in WM transmitters (in which case, the IMD products are transmitted over the air), or in TVBD receivers (in which case the IMD products are only seen by the TVBD).  In either case, the IMD products are nearly indistinguishable from the actual signals. The only way to mitigate this problem is to design receiver systems with a wide dynamic range to avoid operating in a nonlinear region.

Under these uncertainties, it is extremely difficult to abstract features of WM signals and hence develop robust sensing algorithms that meet the suggested guidelines~\cite{She06, SteCor05}. Matched filtering is optimal detection strategy if the TVBD receiver has \textit{a priori} knowledge of the WM signal and the effective channel, which is not practical in general~\cite{Cabric2004}. Energy detection is the simplest sensing algorithm to implement but it is not well suited for WM signals due to two main reasons: 1) it employs non-coherent detection and hence requires longer sensing time to achieve the probability of detection constraints, and 2) the detection thresholds are highly susceptible to the changing noise levels. In fact, it is not possible to meet sensing requirements with a single energy detector under noise uncertainty, even if the sensing is done for a long time~\cite{SahHovTan04,SheTan06,TanSah08}. Another popular method is to estimate the spectral density of the received signal using maximum entropy principle~\cite{Bur67,And74,Jay82,Arc83,CovBook}. While this principle is analytically appealing, it has not been applied yet to WM sensing, perhaps due to the challenges in modeling the practical WM signals accurately. This is an interesting open problem but since modeling is not the main focus of this paper, we leave it as a possible future work. Recently, some interesting approaches for WM signal detection have been proposed in the literature, e.g.,~\cite{Mossa2009} proposes a cyclostationarity-based detector,~\cite{Xu} proposes a Singular Value Decomposition (SVD) based detector,~\cite{Zeng2009} proposes an Eigen value bases detector, and~\cite{Zeng2009a} proposes a covariance based detector. While all these methods perform well under their respective simulation models, their performance in real-world WM sensing is not clear. Moreover most of these methods in their current form are capable of detecting only a single WM signal and the extension of the analysis to multiple WM case does not seem to be trivial.

Motivated by these recent developments, we propose a two-phase SVD based algorithm to detect multiple WM signals. In the training phase, we first compute the average noise autocorrelation matrix from the known sets of noise data, which is then used to whiten the autocorrelation matrices of all the known sets of noise, unmodulated WM signals and modulated signals. These whitened autocorrelation matrices are then used to determine the empirical distributions of the test statistics, which are in turn used to set thresholds for signal detection. In the detection phase, we compute the autocorrelation matrix of the received signal and whiten it by subtracting the average noise autocorrelation matrix computed in the previous phase. The SVD of the resulting autocorrelation matrix is then computed to find the number of WM signals present. Based on the number of signals, the space is decomposed into signal subspace and noise subspace. The autocorrelation of the signals is recovered from the signal subspace, which is then used to find the center frequencies of the WM signals. We distinguish our work from other recent works, especially~\cite{Xu,Zeng2009}, in two major ways. First, we extend previously proposed sub-space methods to the multiple WM sensing case by employing multiple test statistics. Second, we use real signals captured over-the-air to verify the detection performance of the algorithm. We use pass-band signals with $6$ MHz bandwidth to closely emulate the television channel. The experimental data used for verification is representative of harsh channel conditions with received power ranging between -100 to -105 dBm. It is important to note that the problem of differentiating WM signals from other narrow-band signals is not considered in this work and may require techniques from patter recognition or robust statistics for an efficient solution~\cite{KasPoo85, PaRBook}.

The paper is organized as follows. The noise characteristics are briefly studied by using over-the-air signals in Section II. As expected, the noise comes out to be stationary with a near-Gaussian distribution, but is not white. The proposed SVD based two-stage sensing algorithm is discussed in Section III. The detection performance of the proposed algorithm is numerically studied in Section IV. We first study the detection and false alarm probabilities using computer simulation and then verify the working of the algorithm on real over-the-air WM signal data. The paper is concluded in Section V.
\section{Measurement Campaign and Noise Characterization}
For the purpose of noise characterization and algorithm performance evaluation, different data sets were generated containing unmodulated WM signals, modulated WM signals, and background noise respectively. The modulated and unmodulated signal sets had a WM carrier at $8$ MHz, and were sampled at $33$ MHz. The signal was bandpass-filtered to simulate a $6$ MHz television channel. The received power level in these data sets was specified as being between $-100$ dBm and $-105$ dBm. The measurements were captured inside a building and contain a WM transmitter located at varying distances from the receiver. Not only does this provide a very wide range of received signal strength signals, but it also accurately replicates real-world conditions such as multipath and external interference.

\begin{figure}[ht!]
\centering
\includegraphics[width=.75\columnwidth]{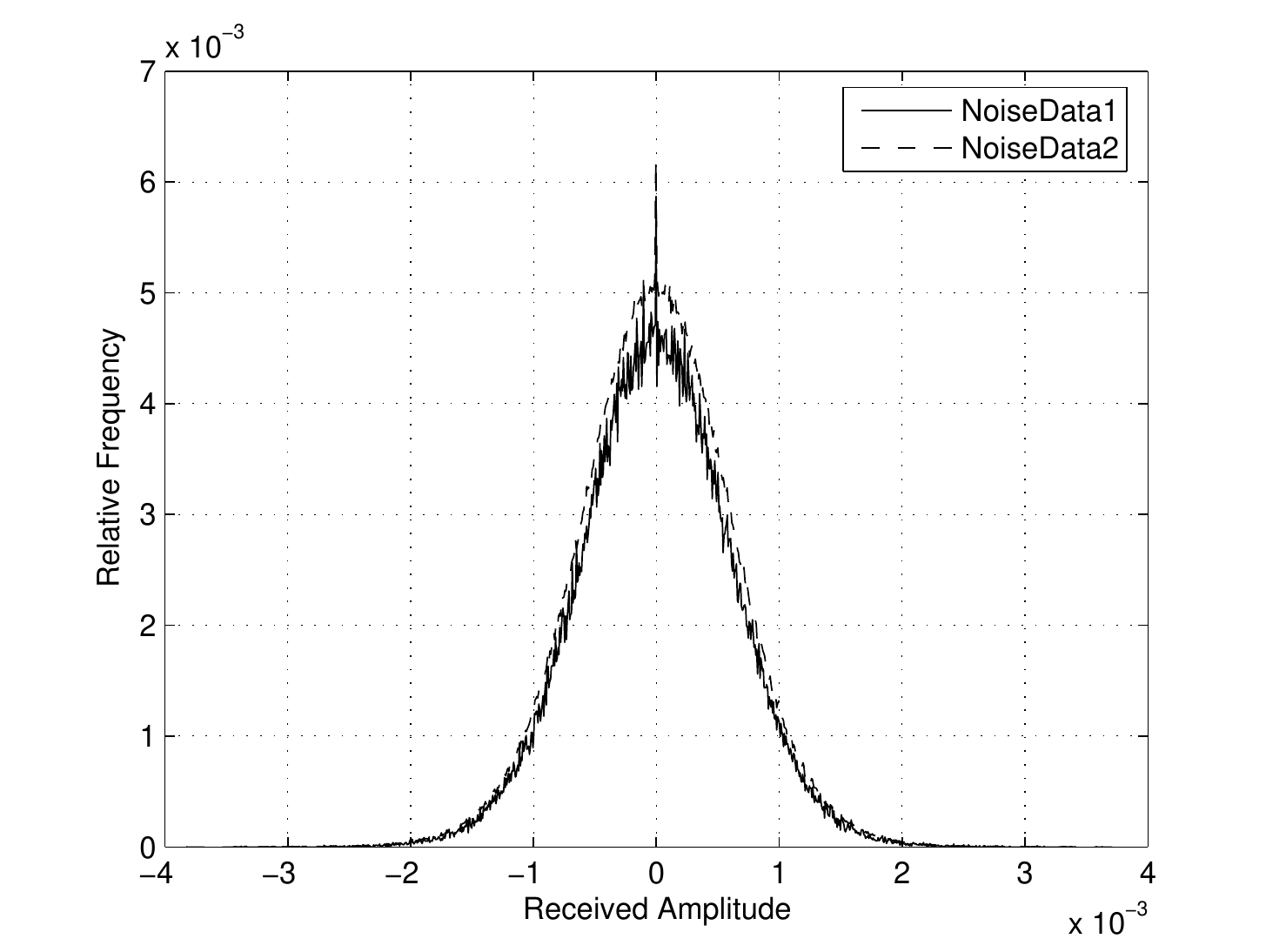}
\caption{Histogram of background noise computed from measurement data.}
\label{noise_histogram}
\end{figure}

\begin{figure}[ht!]
\centering
\includegraphics[width=.75\columnwidth]{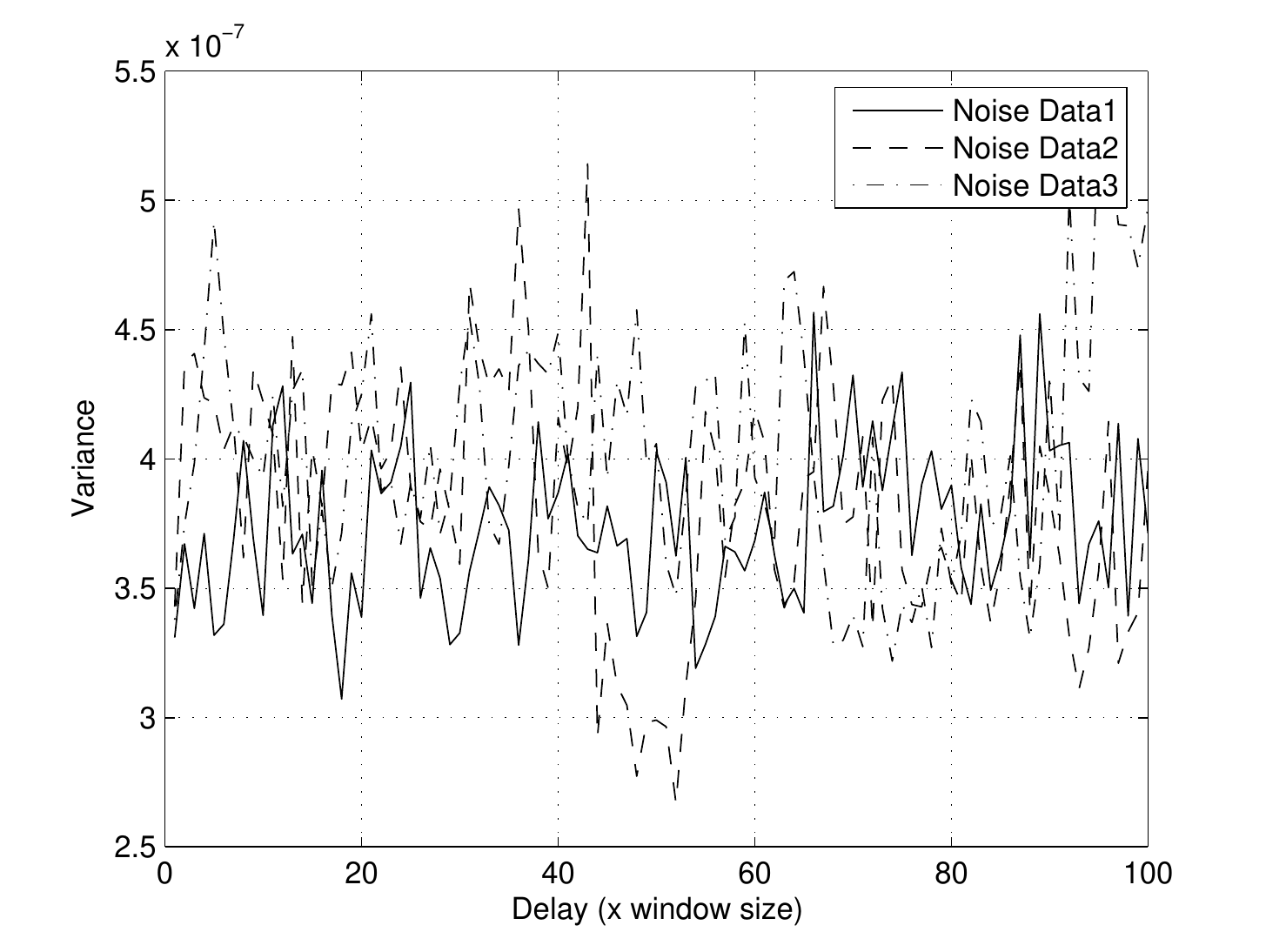}
\caption{Variance of background noise at different time instances.}
\label{noise_meanvar2}
\end{figure}

We use three noise data sets, referred henceforth as \textit{NoiseData\{1:3\}}, to examine various properties of the background noise. The histogram of the noise samples, shown in Fig.~\ref{noise_histogram}, appears to be a truncated Gaussian, although we did not formally test it for Gaussianity. With the assumption that the noise is Gaussian, we can simplify the complexity of the algorithm significantly. The variance of the noise for \textit{NoiseData\{1:3\}} is shown in Fig.~\ref{noise_meanvar2}. The plot indicates that the noise variance does not vary significantly as a function of time. A similar observation was made with the noise mean. The observations are consistent across the noise data sets.

\begin{figure}[t]
\centering
\includegraphics[width=.75\columnwidth]{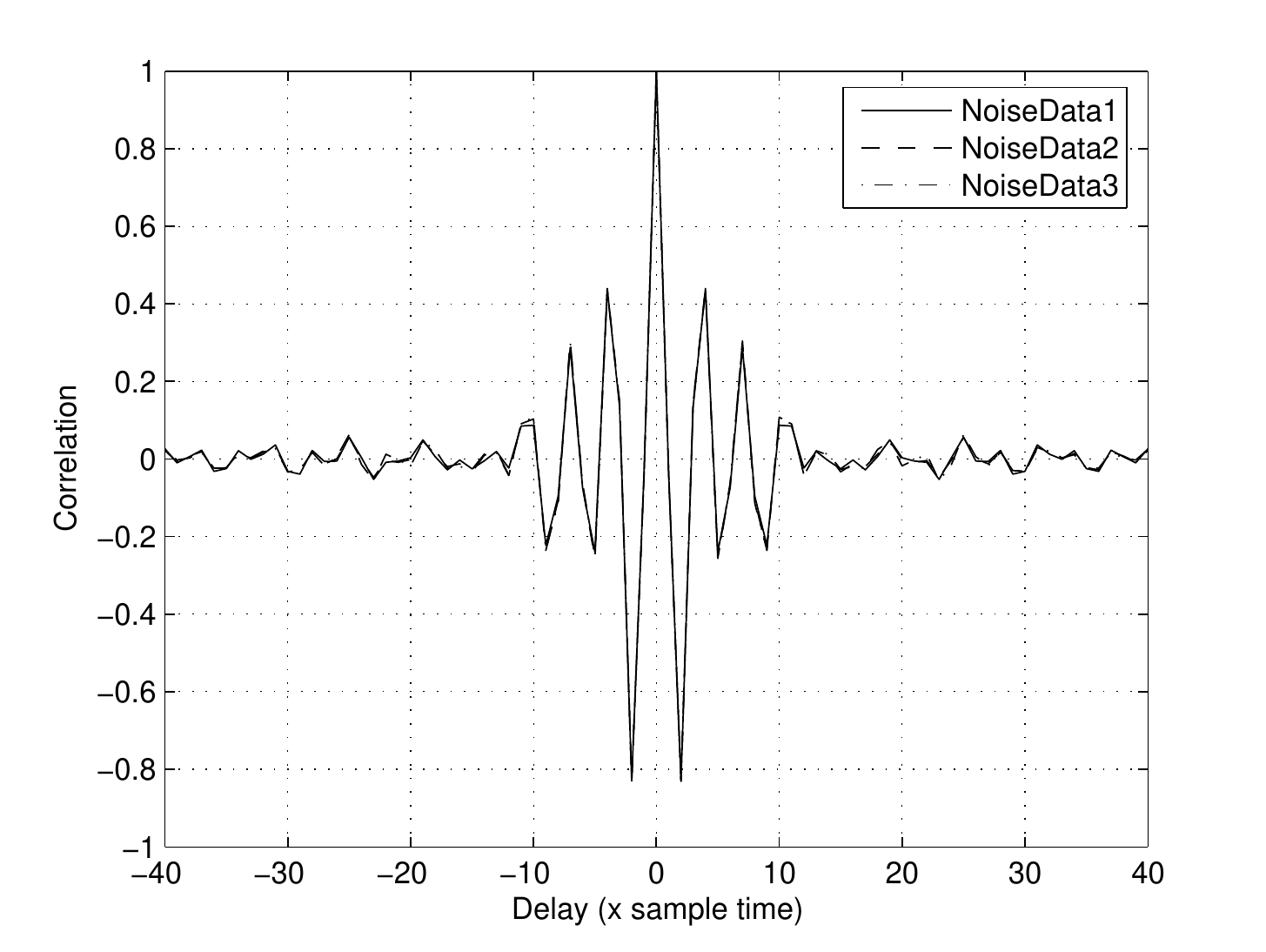}
\caption{Autocorrelation of band-limited background noise samples.}
\label{noise_psdauto1}
\end{figure}

\begin{figure}[t]
\centering
\includegraphics[width=.75\columnwidth]{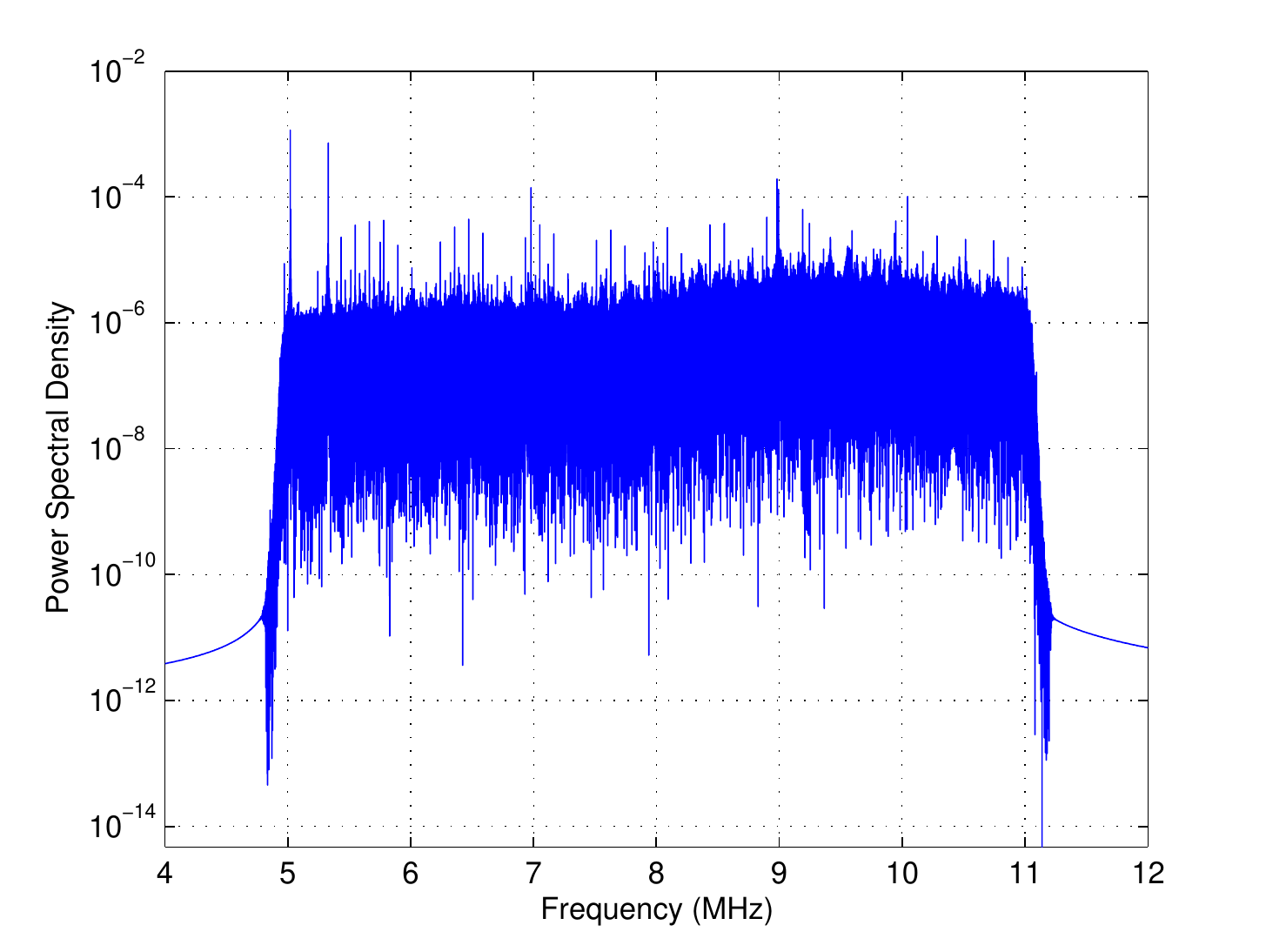}
\caption{PSD of band-limited background noise.}
\label{noise_psdauto2}
\end{figure}

Figs.~\ref{noise_psdauto1} and \ref{noise_psdauto2} show the autocorrelation and power spectral density (PSD) of the noise samples. The PSD clearly shows that the noise was band-pass filtered prior to sampling. The autocorrelation plot reflects this as well, showing correlation up to approximately 15 samples. In addition, the noise correlation properties do not vary significantly across the noise data sets.

\begin{figure}[t]
\centering
\includegraphics[width=.75\columnwidth]{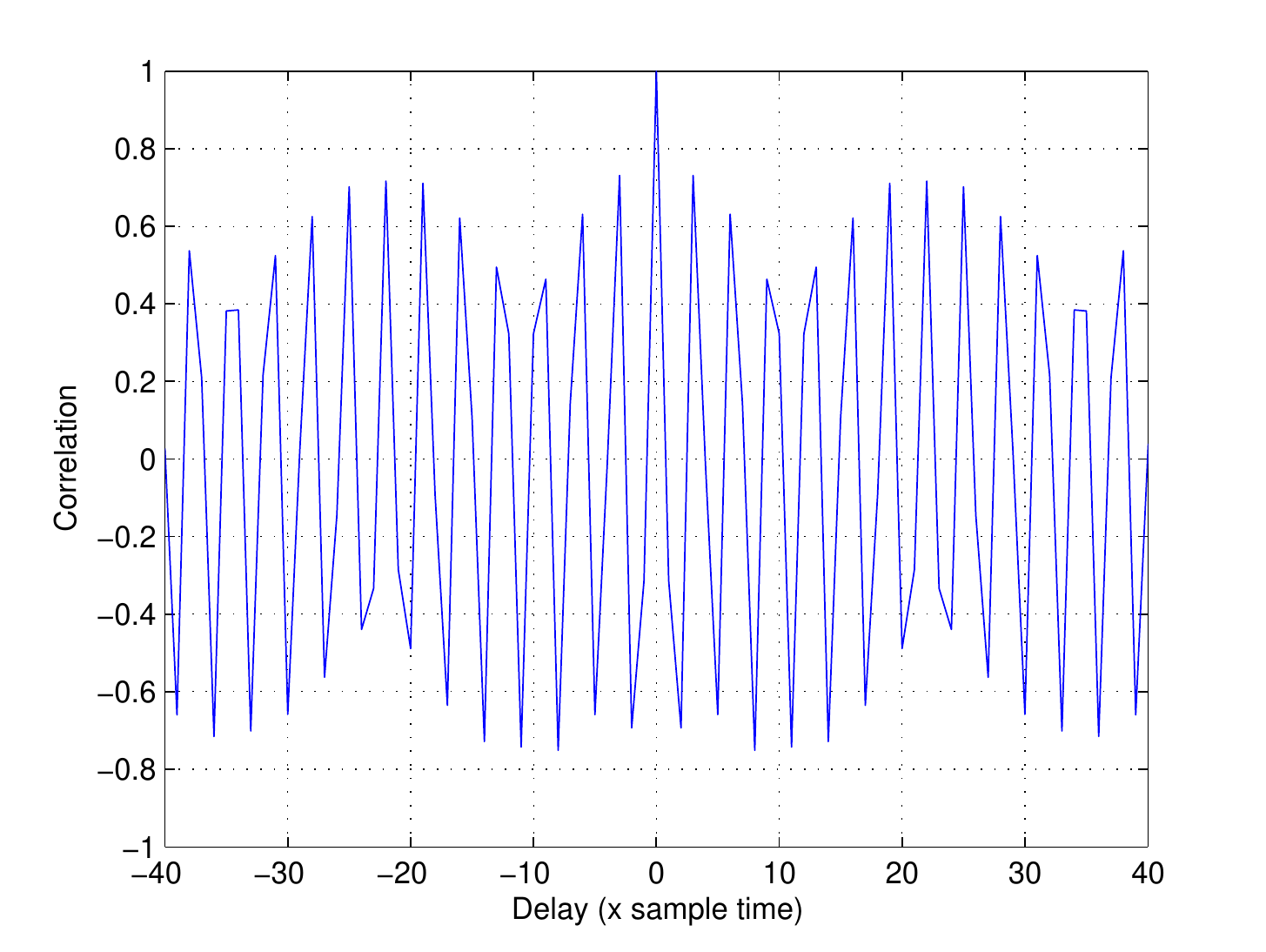}
\caption{Autocorrelation of loud (modulated) WM signal with colored noise.}
\label{signal_psdauto1b}
\end{figure}

\begin{figure}[t]
\centering
\includegraphics[width=.75\columnwidth]{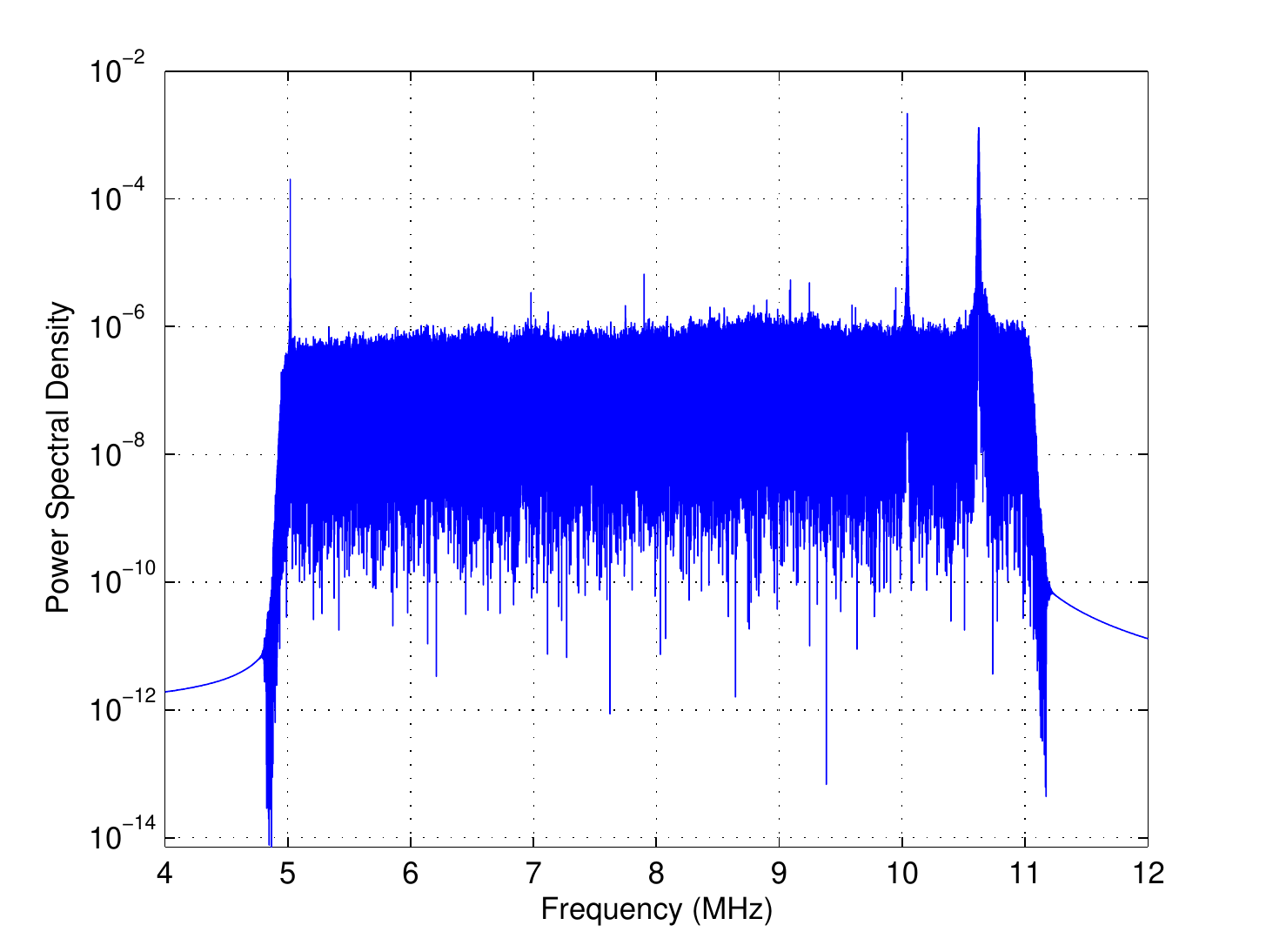}
\caption{PSD of modulated WM signal with colored noise.}
\label{signal_psdauto2}
\end{figure}

Figs.~\ref{signal_psdauto1b} and \ref{signal_psdauto2} show the autocorrelation and PSD of the data samples containing a modulated WM signal. As expected, the autocorrelation differs significantly from Fig.~\ref{noise_psdauto1} in that the correlation with signal present extends out far past 15 samples. It is this property that allows us to use a correlation-based technique to determine whether a WM signal is present in a given data set or not. The autocorrelation of the data samples containing a silent WM signal resembles that in Figure~\ref{signal_psdauto1b}, with the exception that the correlation is lesser.

These observations are in line with the intuition and we conclude that the noise is stationary with a near-Gaussian distribution, but is not white. Rather, the noise is band-limited (colored) due to filtering and sampling in the receiver.

\section{SVD Based WM Sensing Algorithm}
The working of our proposed algorithm can be classified into two main phases, viz., training phase and detection phase. Main steps involved in these two phases are as follows:
\subsection*{Training Phase:}
\begin{enumerate}
\item Characterize the noise by using the training data sets \textit{NoiseData\{1:3\}} containing no WM signal.
\item Compute the average noise autocorrelation matrix from all the training data sets.
\item Subtract this average noise autocorrelation matrix from the autocorrelation matrix of each \textit{NoiseData} to get the \textit{whitened noise correlation matrices}.
\item Determine the thresholds of the test statistics by using training sets of noise and WM signal.
\end{enumerate}
\subsection*{Detection Phase:}
\begin{enumerate}
\item Compute the autocorrelation matrix of the received signal.
\item Whiten the noise by subtracting the average noise autocorrelation matrix from the received signal correlation matrix.
\item Find the SVD of the resulting correlation matrix and determine the number of signals present by comparing the test statistics with the thresholds.
\item Decompose the space into signal sub-space and noise sub-space according to the number of WM signal present.
\item Recover the autocorrelation of the signal from the signal autocorrelation matrix (signal sub-space).
\item Find the center frequencies by the Fourier analysis of the signal autocorrelation.
\end{enumerate}

With this brief introduction to the proposed algorithm, we now discuss it in detail. The received signal samples are modeled as:
\begin{equation}
x(n) = s(n) + \eta(n),
\end{equation}
where $s(n)$ represents the transmit signal and $\eta(n)$ models the cumulative effect of thermal noise and interference, referred henceforth as only noise. For a received signal sequence of length $L$, the $L\times L$ autocorrelation matrix can be represented as $\textbf{R}_{x} = \textbf{R}_{s} + \textbf{R}_{\eta}$, where $\textbf{R}_{x} = E\left\{ x(n) \; x^{H}(n)  \right\}$ is the autocorrelation matrix of the received samples, $\textbf{R}_{s} = E\left\{ s(n) \; s^{H}(n)  \right\}$ is the autocorrelation matrix of the transmitted signal samples, and $\textbf{R}_{\eta} = E\left\{ \eta(n) \; \eta^{H}(n) \right\}$ is the autocorrelation matrix of the noise samples. Note that $s^{H}$ denotes the conjugate transpose of vector $s$. In the case when the noise is white, $\textbf{R}_{\eta} = \sigma^{2}_{\eta} \; \textbf{I}_{L}$, where $\textbf{I}_{L}$ is an identity matrix of order $L$. The value of $L$ is chosen as a trade-off between complexity and performance. We will comment more on the choice of $L$ in the numerical results section.

\subsection{Noise Whitening}
To perform noise whitening, we first compute an estimate of the average noise autocorrelation matrix $\widehat{\textbf{R}}_{\eta}$ by taking average of the autocorrelation matrices of the three noise sets \textit{NoiseData\{1:3\}}. This average autocorrelation matrix is then used to cancel off the noise correlation from the received signal autocorrelation matrix as follows:
\begin{eqnarray}
\label{eqn:noisewhite}
\widehat{\textbf{R}}_{s} = \textbf{R}_{x} - \widehat{\textbf{R}}_{\eta} = \textbf{R}_{s} + \textbf{R}_{\mu},
\end{eqnarray}
where $\textbf{R}_{\mu}$ is the correlation matrix of the residual noise. $\widehat{\textbf{R}}_{s}$ is the estimate of the correlation matrix of the WM signal assuming that the correlation matrix $\textbf{R}_{\mu}$ of the residual noise is negligible compared to $\textbf{R}_{s}$. This basic method should not be confused with the popular noise whitening methods where the correlated noise is passed through a filter to achieve zero mean and identity covariance matrix. There are two main reasons for choosing this basic method, despite it being not as accurate as the standard whitening method: 1) it is difficult to design the whitening matrix with the limited noise data sets, and 2) more importantly, it serves the purpose of removing the prominent interference signals that might be present in the ambience.

\subsection{SVD and Test Statistic Threshold}

The SVD method can be applied to the cleaned autocorrelation matrix of received signal $\widehat{\textbf{R}}_{s}$ to obtain
\begin{eqnarray}
\label{eqn:svd}
\widehat{\textbf{R}}_{s} = \textbf{U} \; \textbf{S} \; \textbf{V}^{H} = [\textbf{U}_{s} \textbf{U}_{\mu}]
\left[
\begin{array}{c c}
	\textbf{S}_{s} &  0 \\
	0 & \textbf{S}_{\mu}
\end{array} \right]
[\textbf{V}_{s} \textbf{V}_{\mu}]^{H},
\end{eqnarray}
where $\textbf{S}_{s}$ and $\textbf{S}_{\mu}$ are diagonal matrices whose values correspond to the singular values in the signal subspace and noise subspace, respectively. It is straightforward to show that each WM signal produces two non-zero singular values. Therefore, for $N_s$ WM signals, $diag(\textbf{S}_{s})$ will be $[\lambda_{1},\lambda_{2}, \ldots \lambda_{2 N_{s}}] $ and $diag(\textbf{S}_{\mu})$ will be $[\lambda_{2 N_{s} + 1},\lambda_{2 N_{s} + 2}, \ldots \lambda_{L}]$, where the singular values are arranged in the decreasing order. E.g., if there is only one WM signal present, $diag(\textbf{S}_{s}) = [\lambda_{1},\lambda_{2}]$ and $diag(\textbf{S}_{u}) = [\lambda_{3},\lambda_{4} \ldots \lambda_{L}]$. It should be noted that the singular values corresponding to noise sub-space are all of same order and much smaller than those of the signal sub-space. The natural test statistic, therefore, is to take the ratio of alternate singular values, i.e., $ \lambda_{2 x -1} / \lambda_{2 x +1}$, $1 \le x \le (L-2)/2$. In case only white noise is present, we get $\lambda_{2 x -1} / \lambda_{2 x +1} \cong 1$, $\forall\ x$. However, if $N_{s}$ number of signals are present, we get $\lambda_{2 x -1} / \lambda_{2 x +1} >> 1$ for $1 \le x \le N_s$ and $\lambda_{2 x -1} / \lambda_{2 x +1} \cong 1$ for $N_s+1 \le x \le (L-2)/2$.

The threshold values, $\lambda_{\tau_i}$ for $1 \le i \le (L-2)/2$, are determined by applying SVD to the whitened autocorrelation matrices of known sets of noise, unmodulated WM signals and modulated WM signals. The empirical distributions of test statistics are determined in each of these cases and thresholds are set to achieve predetermined probability of false alarm ($P_{fa}$) or probability of detection ($P_{d}$). Those familiar with the problem of determining empirical distributions are likely to immediately recognize that it is very hard to get a clean estimate from a very limited set of observations. Therefore, we identify the duration and frequency of training as two main practical parameters that affect the performance of the proposed algorithm.

\subsection{Center Frequencies of WM Signals}

After identifying the number of WM signals present, the space is decomposed into noise and signal subspace as shown in equation~(\ref{eqn:svd}). The autocorrelation of the cleaned signal is recovered, which is nothing but the first row of the autocorrelation matrix $\textbf{R}_{s}$. PSD is then determined by taking the Fourier transform of the signal autocorrelation. The locations of the WM signals can be estimated by the locations of $N_s$ highest peaks in the PSD. Since the autocorrelation matrix $\textbf{R}_{s}$ is reconstructed only from the signal subspace, the resulting PSD is much cleaner than that obtained from correlation matrix $\textbf{R}_{x}$ or even $\widehat{\textbf{R}}_{s}$. This makes determining the location of the WM signals much easier and more robust.

\section{Numerical Results}
In this section, we determine the detection performance first by using computer simulation and then by applying the algorithm on real over-the-air WM signals. The computer simulation is used to study finer performance details, which is not always possible to do in case of a limited set of real-world data.

\subsection{Computer Simulation}
We generate WM signals based on the procedure explained in \cite{Clanton2007}. Two different data sets are generated: the training set and the test set. The two data sets essentially consist of same type of data. The training set is used to obtain the thresholds for the test statistics, while the test set is used to measure the performance of the algorithm. The sampling rate used for the simulation is $F_s = 33.33$ MHz and each test set is $20000$ samples long.

\subsubsection{Generation of Training and Test Set}
\begin{figure}[t]
	\centering
	\includegraphics[width=.75\columnwidth]{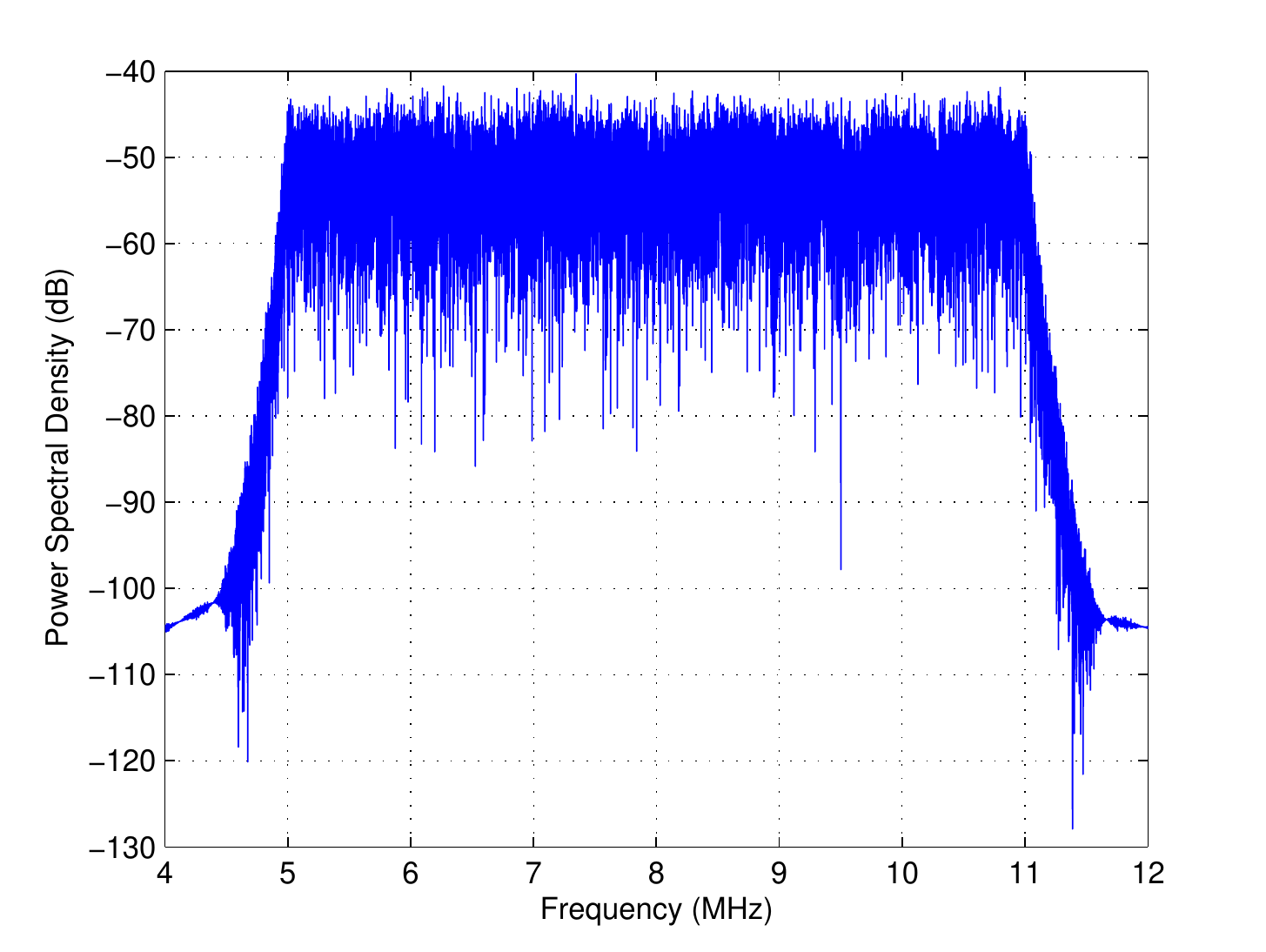}
	\caption{PSD of simulated band-limited background noise.}
	\label{fig:psd_sim_noise}
\end{figure}

\begin{figure}[t]
	\centering
	\includegraphics[width=.75\columnwidth]{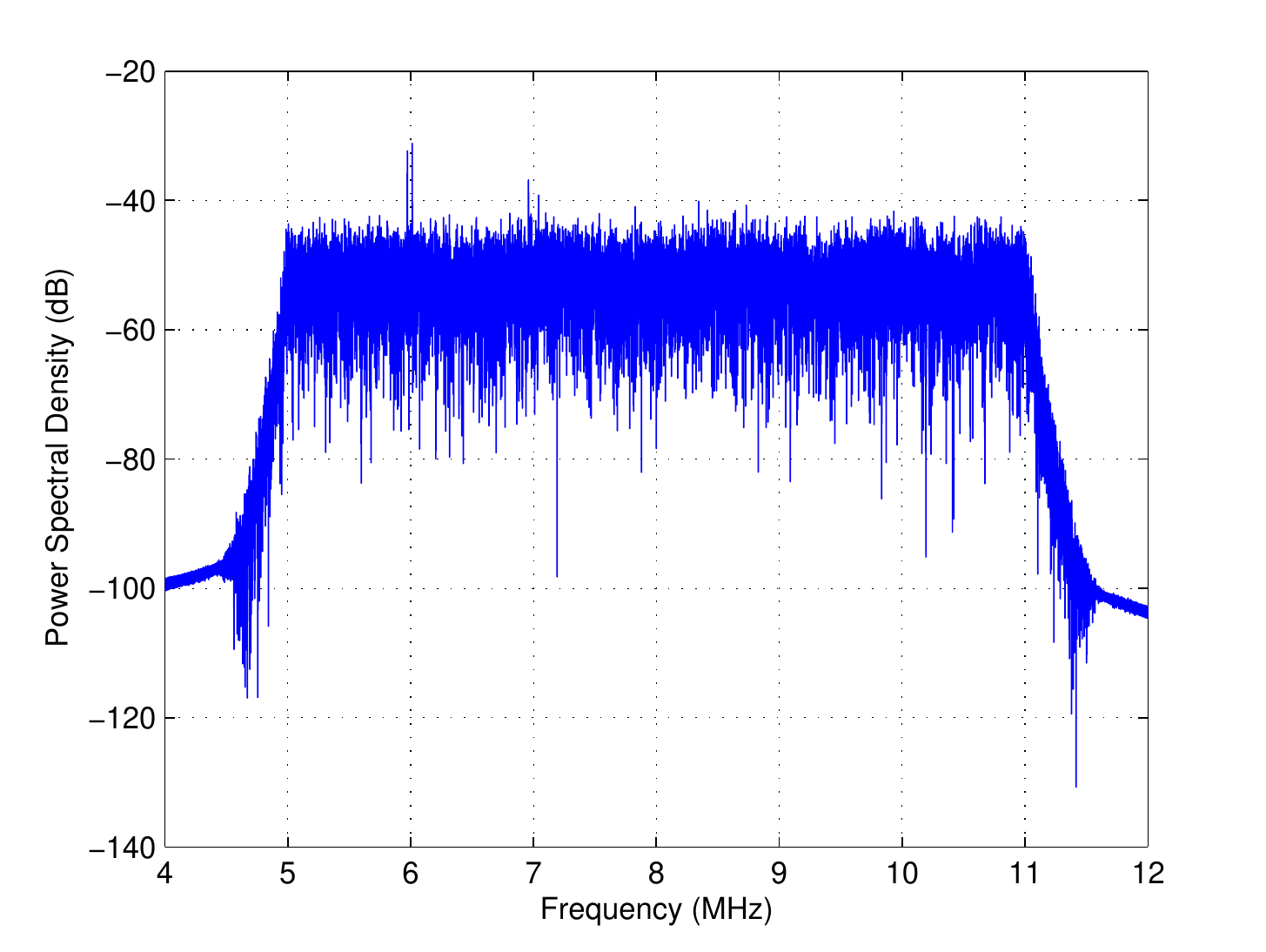}
	\caption{PSD of simulated WM signal in loud mode with colored noise.}
	\label{fig:psd_sim_wmandnoise}
\end{figure}

\begin{figure}[t]
	\centering
	\includegraphics[width=.75\columnwidth]{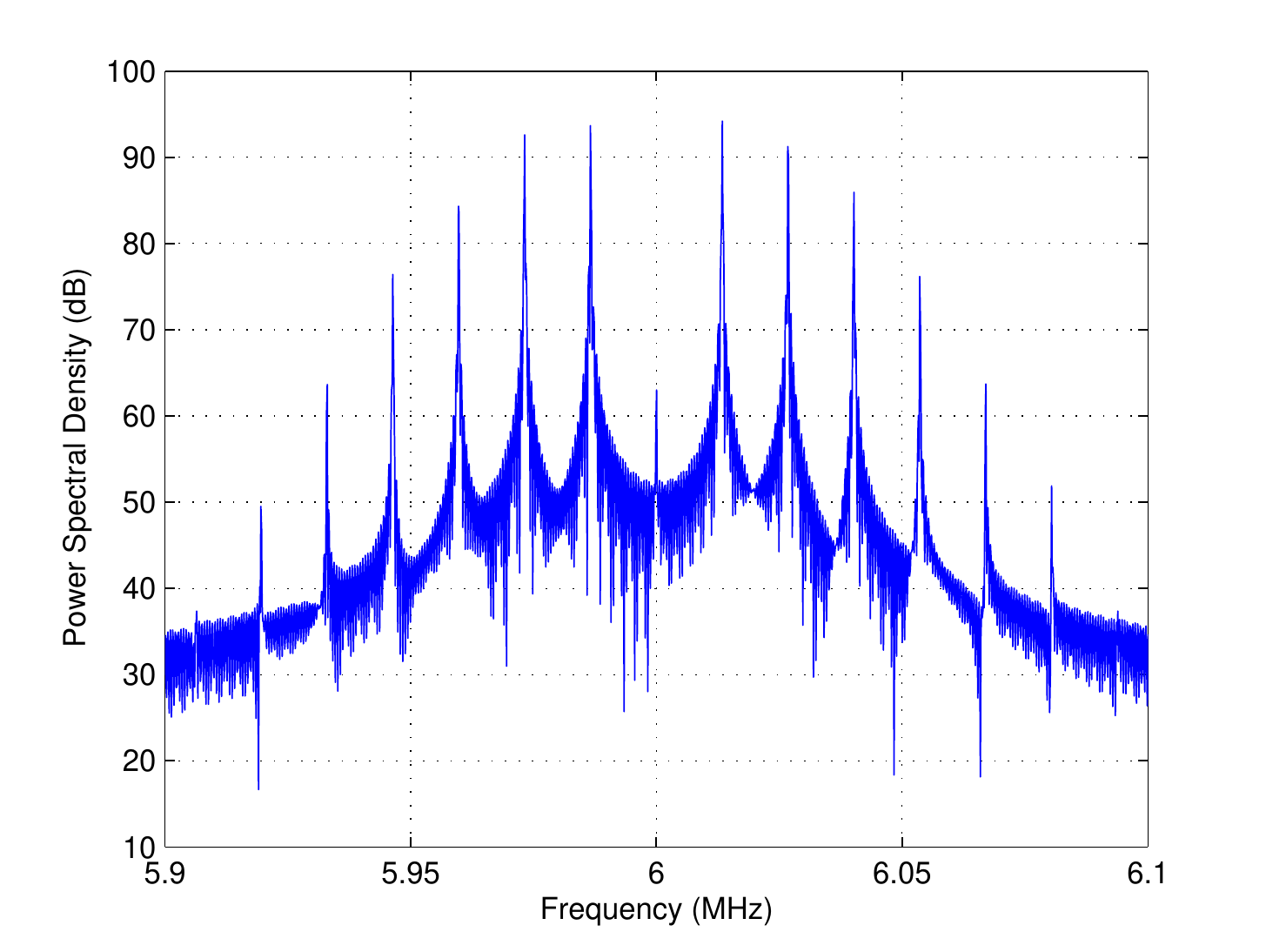}
	\caption{PSD of simulated wireless microphone signal: loud speaker.}
	\label{fig:psd_sim_loudwm}
\end{figure}

We generate both the noise-only data and the WM signals embedded in noise for both the training and the test sets. The noise signal is generated by filtering additive white gaussian noise (AWGN) with a bandpass filter of $6$ MHz bandwidth centered at the IF of $8$ MHz. Fig.~\ref{fig:psd_sim_noise} presents the PSD of noise-only signal. For data with WM signals, the number of signals is randomly varied from one to five. The locations of the signals were fixed to pre-specified frequencies of $6 , 7 , 8 , 9$ and $10$ MHz for ease of verification. For example, if the number of signals was selected to be two, the signals would be fixed at the frequencies of $6$ and $7$ MHz. If five were present, they would be centered at $6, 7, 8, 9$ and $10$ MHz. The SNR of the WM signal is varied from -$30$ dB to -$15$ dB. For this paper, only the WM signal with loud speaker is simulated. The silence and soft speaker modes are left as a future work. Fig.~\ref{fig:psd_sim_wmandnoise} shows the PSD of WM signals at $6$ and $7$ MHz at SNR of -$20$dB. Fig.~\ref{fig:psd_sim_loudwm} shows the PSD of wireless microphone signal with loud speaker that was used for the simulation.

\subsubsection{Test Statistic Threshold Determination}
\begin{figure}[t]
	\centering
	\includegraphics[width=.60\columnwidth]{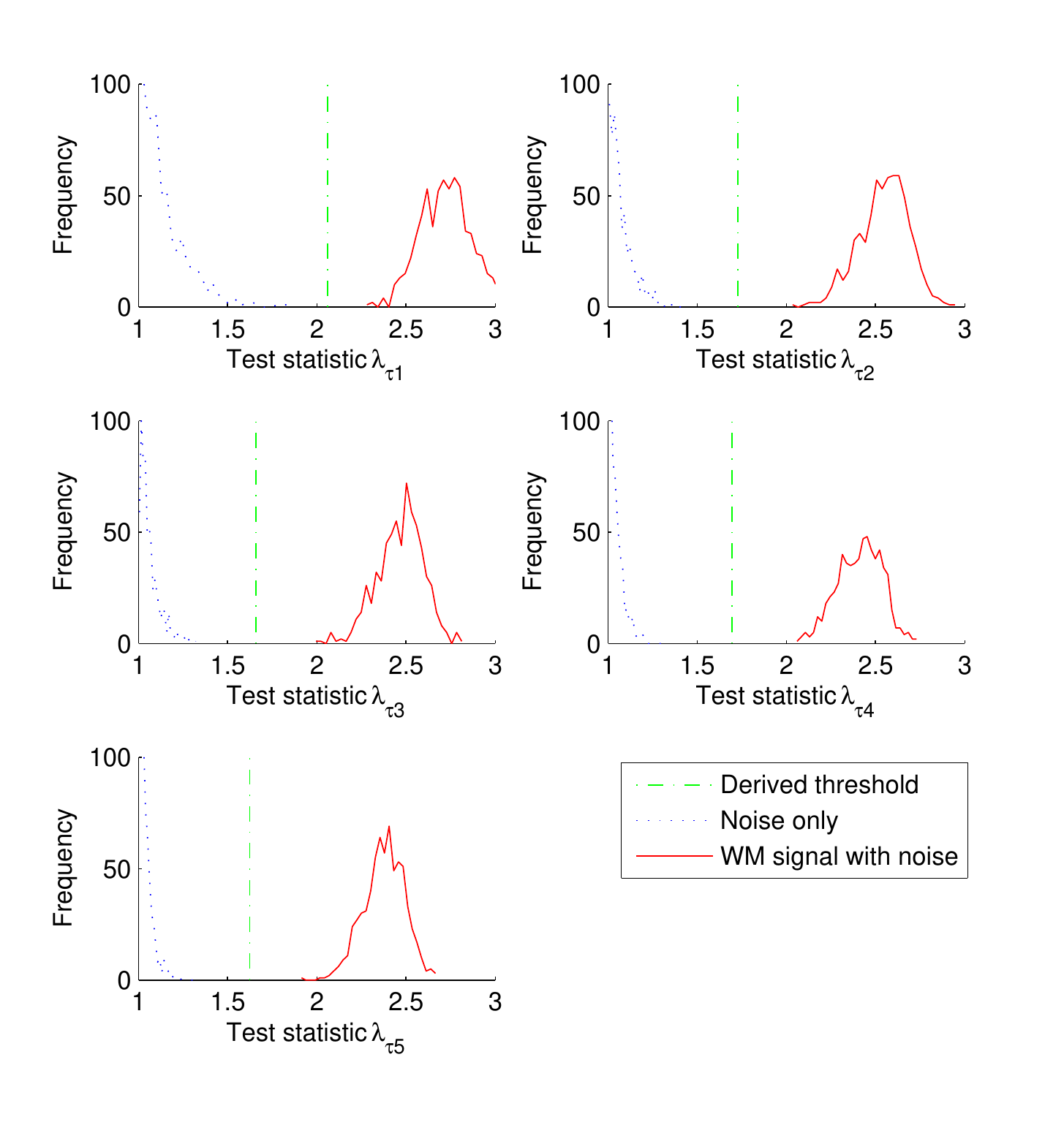}
	\caption{Histogram of test statistic at SNR $=-15$ dB.}
	\label{fig:distribution_-15dB}
\end{figure}

\begin{figure}[t]
	\centering
	\includegraphics[width=.60\columnwidth]{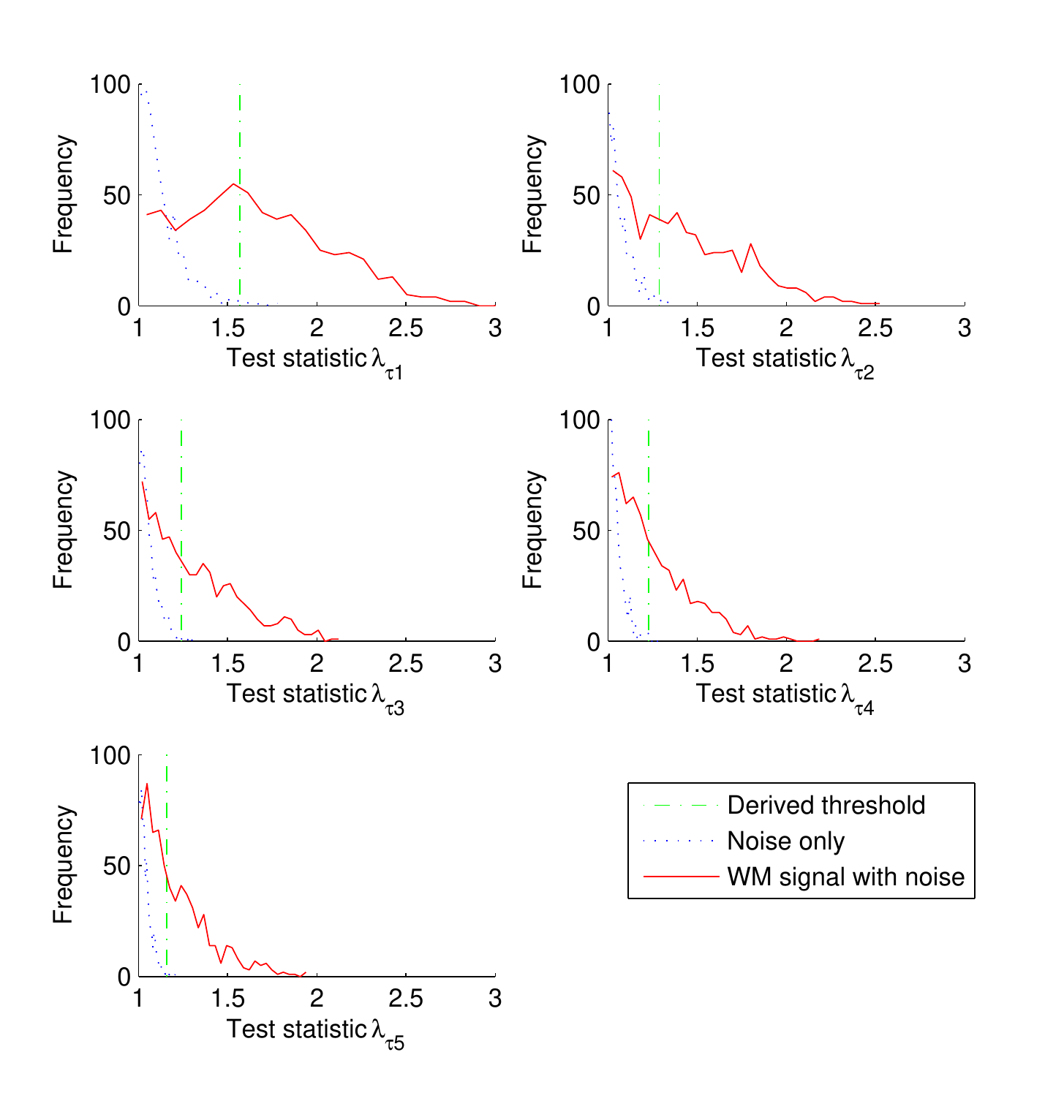}
	\caption{Histogram of test statistic at SNR $=-25$ dB.}
	\label{fig:distribution_-25dB}
\end{figure}

In order to derive the threshold for the test statistics, the distribution of the test statistics was first obtained. As shown in Fig.~\ref{fig:distribution_-15dB}, at relatively high SNR of -15dB, the distributions of test statistics for noise-only and signal plus noise cases are well separated. However, Fig.~\ref{fig:distribution_-25dB} shows that at relatively low SNR of -25 dB, the distribution of test statistics of the two cases are not well separated, which is inline with the intuition. The thresholds are determined using simple hypothesis testing.
\[
\begin{array}{cl}
	H_{0}, & \mbox{No wireless microphone present} \\
	H_{1}, & \mbox{Wireless microphones present}
\end{array}\
\]
Given the distributions of the test statistics, we find $\lambda_{\tau}$ such that $P_{fa} = P[\lambda > \lambda_{\tau} | H_{0}] \leq 0.1 $, $\lambda$ is the test statistic, and $\lambda_{\tau}$ is the threshold. However, since the distributions do not perfectly match the true probability distribution of the test statistics, the threshold was later adjusted iteratively by finding the actual $P_{fa}$. Five thresholds were determined for each SNR value ranging from -30 dB to -15 dB. Table~\ref{table:thresholds} shows the thresholds obtained for each SNR.

\subsubsection{Detection Results}
\begin{figure}[t]
	\centering
	\includegraphics[width=.75\columnwidth]{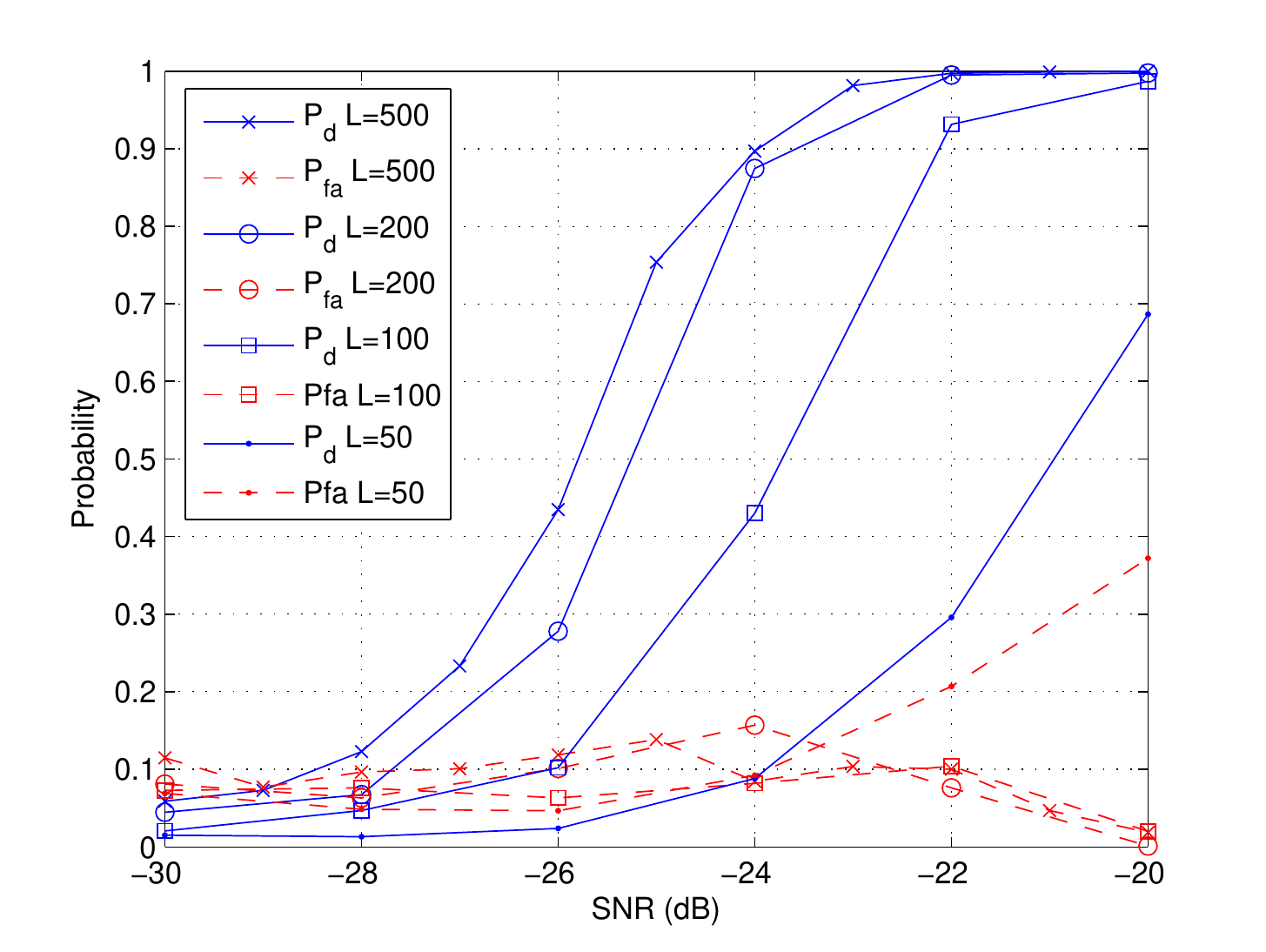}
	\caption{Detection result for simulated data in AWGN.}
	\label{fig:result_sim_pd_pfa}
\end{figure}

\begin{figure}[t]
	\centering
	\includegraphics[width=.70\columnwidth]{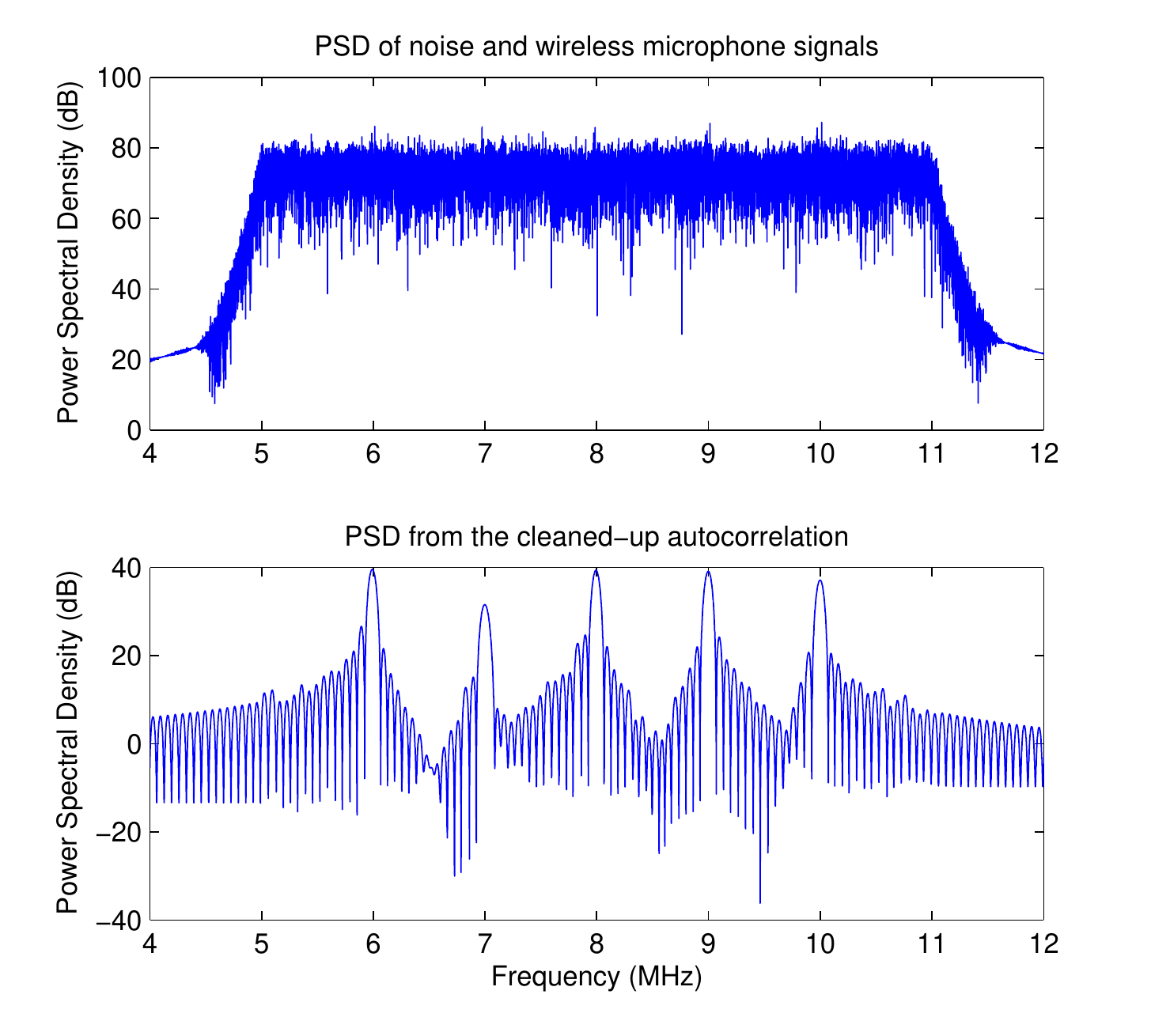}
	\caption{PSDs of the original data and the cleaned-up autocorrelation: five signals present.}
	\label{fig:PSD_results}
\end{figure}

Fig.~\ref{fig:result_sim_pd_pfa} shows the probability of detection and probability of false alarm based on the derived thresholds for AWGN case. \textit{The detection results for the Rayleigh fading case will be included in the future version of this paper.} The results show that the algorithm successfully detects multiple wireless microphone signals with high probability at SNR as low as -23 dB if a sufficiently long signal sequence is taken. Although not perfectly, the probability of false alarm is also approximately bounded by 0.1. Fig.~\ref{fig:PSD_results} shows the PSD of the original signal  at SNR of -25dB and the PSD obtained after processing the original signal. It shows that five signals at $6, 7, 8, 9$ and $10$ MHz are correctly detected.

\subsubsection{Effect of Varying $L$}
Calculating the autocorrelation matrix, which is one of the steps required in the algorithm, is computationally expensive for longer sequences (higher values of $L$). Furthermore, working with large matrices is typically not desirable for real-time sensing algorithms. Throughout the paper, we have considered the value of $L$ to be $500$, which was chosen to make sure that the resulting signal sequence was long enough to achieve the best possible detection performance. We now vary the value of $L$ and study the trade-off between performance and complexity in Fig.~\ref{fig:result_sim_pd_pfa}. As expected, the decreasing value of $L$ degrades the detection performance of the algorithm. We note that $L\approx 200$ is a good tradeoff between performance and complexity.

\subsection{Using Real Data}
\begin{figure}[t]
 	\centering
	\includegraphics[width=.75\columnwidth]{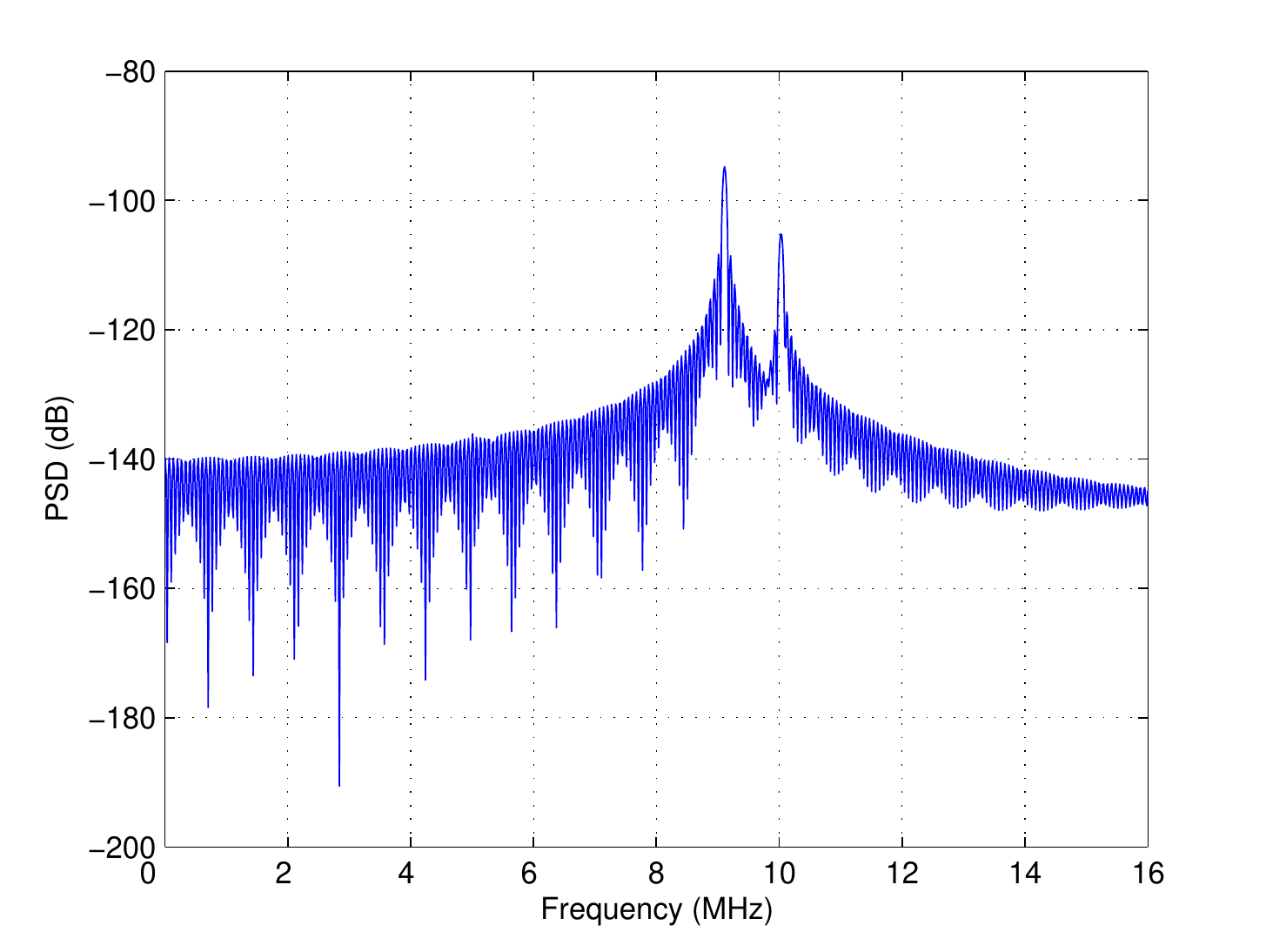}
	\caption{PSDs obtained from the cleaned-up autocorrelation function $\textbf{R}_{s}$.}
	\label{fig:psd_loc2}
\end{figure}

The proposed algorithm has been verified and tested on real over-the-air data. The empirical thresholds that were used for detection are as follows: $\lambda_{\tau_{1}} = 1.8741$, $\lambda_{\tau_{2}} = 1.4505$, $\lambda_{\tau_{3}} = 1.5743$, $\lambda_{\tau_{4}} = 1.4806$ and $\lambda_{\tau_{5}} = 1.3152$. Fig.~\ref{fig:psd_loc2} shows the PSD obtained from the cleaned-up autocorrelation function $\textbf{R}_{s}$ for a data set containing WM signal. As mentioned earlier, it is not practical in this case to estimate detection and false alarm probabilities due to the limited data set. Therefore, we analyze the data sets by visual inspection as an extra step independent of the proposed algorithm. Interestingly, the visual inspection has revealed that a few data sets contain more number of signals than are detected by the proposed algorithm. Technically this means that some of these signals are not leading to independent columns of the autocorrelation matrix, and hence the number of non-zero singular values is lower than expected. One possibility is that there could be intermodulation products present in the received signal that share the same subspace as the WM signals itself, in which case the SVD algorithm may not observe any singular values corresponding to the intermodulation products.

\section{Conclusion and Future Direction}
Detecting WM signals in poor SNR conditions using test equipment is a difficult task. Doing the same task with a consumer-grade receiver over a wide band will prove to be even more challenging. This paper has shown how a simple sub-space method can detect the presence of extremely weak WM signals and determine what frequencies they operate on.

The current version of the proposed algorithm faces one drawback. It is not capable of examining each signal individually to determine whether it exhibits the characteristics of a frequency-modulated analog audio signal. The next step in this process is to incorporate this information into the algorithm, as well as the capability to calculate potential IM3 products and compare the results to the detected signals to reduce the possibility of false positives (especially over multiple TV channels). More measurement campaigns have to be conducted under different wireless environments in order to obtain more statistically accurate calculations of the test statistic thresholds. One other important research direction would be to evaluate the complexity and performance of the proposed algorithm on software radio platforms including small form factor handsets.

\section*{Acknowledgments}
The authors wish to thank Qualcomm Inc., particularly Dr. Steve Shellhammer, for providing us this research opportunity. The wireless microphone signals used in evaluating the spectrum sensing techniques described in this paper were provided by Qualcomm as part of the Qualcomm Cognitive Radio Contest. The contest was held between students from a number of universities in North America to find out which team had the best wireless microphone spectrum sensing technique.

\bibliographystyle{IEEETran}
\bibliography{Springer_WirelessMic_ArXiv_V1}

\begin{table}[h!b!p!]
\centering
\caption{Test Statistic Thresholds}
\begin{tabular}{| c || r | r | r | r | r |}
\hline
\hline SNR (dB) & $\lambda_{\tau1}$ & $\lambda_{\tau2}$ & $\lambda_{\tau3}$ & $\lambda_{\tau4}$ & $\lambda_{\tau5}$ \\ \hline
\hline
		-30 & 1.568	& 1.252	& 1.251	& 1.196	& 1.143 \\ \hline
		-29 & 1.655	& 1.294	& 1.221	& 1.194	& 1.174 \\ \hline
		-28 & 1.505	& 1.357	& 1.205	& 1.202	& 1.168 \\ \hline
		-27 & 1.657	& 1.290	& 1.236	& 1.198	& 1.205 \\ \hline
		-26 & 1.459	& 1.304	& 1.260	& 1.166	& 1.217 \\ \hline
		-25 & 1.569	& 1.288	& 1.242	& 1.228	& 1.157 \\ \hline
		-24 & 1.463	& 1.387	& 1.229	& 1.231	& 1.227 \\ \hline
		-23 & 1.678	& 1.319	& 1.220	& 1.205	& 1.176 \\ \hline
		-22& 1.517	& 1.273	& 1.252	& 1.198	& 1.158 \\ \hline
		-21 & 1.724	& 1.538	& 1.342	& 1.252	& 1.223 \\ \hline
		-20 & 1.806	& 1.528	& 1.361	& 1.359	& 1.322 \\ \hline
		-19 to -15 & 2.060	& 1.726	& 1.658	& 1.694	& 1.623 \\ \hline
\hline
\end{tabular}
\label{table:thresholds}
\end{table}

\end{document}